\documentclass[useAMS,usenatbib]{mn2e}

\usepackage{psfig,subfigure}

\newcommand{\bvec}[1]{\mbox{\boldmath ${#1}$}}

\begin{document}
%\topmargin 3.5cm
 
%\thesaurus{07(08.01.2; 08.13.1; 08.09.1; 08.01.3; 02.12.3; 02.16.2)}

\title[Polar fields for AB Doradus]
{Polar fields for AB Doradus}

\author
[T. McIvor]
{T. McIvor$^1$, M. Jardine$^1$, A. Collier Cameron$^1$ K.Wood$^1$ and J.-F. 
Donati$^2$
\thanks{E-mail: tm29@st-and.ac.uk}
\\
$^1$School of Physics and Astronomy, Univ.\ of St~Andrews,
St~Andrews,
Scotland KY16 9SS \\
$^2$Laboratoire d$^1$Astrophysique,Observatoire Midi-Pyr\'en\'ees, 14
Av. E. Belin, F-31400 Toulouse, France \\
\\
}

\date{Received; accepted }

\maketitle

\begin{abstract}
    
Polar spots are often observed on rapidly-rotating cool stars, but the
nature of the magnetic field in these spots is as yet unknown.  While
Zeeman-Doppler imaging can provide surface magnetic field maps over
much of the observed stellar surface, the Zeeman signature is
suppressed in the dark polar regions.  We have determined the effect
on the global coronal structure of three current models for this polar
field: that it is composed (a) of unipolar field, (b) of bipolar regions or (c) 
of
nested rings of opposite polarity.  We take as an example the young,
rapid rotator AB Dor(P$_{\rm rot}$ = 0.514 days).  By adding these model
polar fields into the surface field maps determined from Zeeman-Doppler
imaging, we have compared the resulting coronal structure with the
observable properties of the corona - the magnitude and rotational
modulation of the X-ray emission measure and the presence of slingshot
prominences trapped in the corona around the Keplerian co-rotation
radius. We find that only the presence of a unipolar spot has any 
significant effect on the overall coronal structure, forcing much of 
the polar field to be open. 

\end{abstract}


\section{Introduction}

Since the advent of Doppler imaging it has been possible to map the
distributions of starspots on rapidly-rotating stars.  This has
revealed that starspots are typically found not only at the low
latitudes where sunspots emerge, but also at very high latitudes,
extending all the way to the pole \citep{1996IAUS..176..289S}.  At
present there is no consensus on the origins of these high-latitude
spots.  The only thing that is clear is that rapid rotation seems to
play an important role.

Three possible models have been proposed to explain the latitudinal
distribution of starspots.  As early as 1992, \citet{1992A&A...264L..13S}
proposed that flux tubes formed deep in the stellar convective
zone would be deflected poleward by Coriolis forces as they attempted
to rise buoyantly to the surface \citep{2000A&A...355.1087G}.
This would result in a polar cap that was formed of mixed polarity
regions of a similar nature to those at lower latitudes.  More
recently, \citet{2001ApJ...551.1099S} have presented a model for flux
emergence on rapid rotators.  A poleward meridional flow carries
bipoles towards the poles while at the same time diffusion attempts to
disperse them.  The inclination of these bipoles to the equator causes
the trailing polarities to reach the pole first, leading to a flux
distribution which resembles concentric rings of alternating polarity
encircling the pole.  Finally, \citet{2002MNRAS.329..102K} have suggested on
the basis of radio observations that the polar caps are formed of
unipolar field, perhaps part of a large-scale dipole.

In all of these scenarios, the polar caps would appear dark in a
Doppler image if the polar fields were strong enough to suppress
convection.  What these Doppler images cannot reveal is whether the
polar field is composed of closed loop structures which may be bright
in X-rays, or open field regions which may contribute to the sellar
wind.  This question becomes of importance when considering the
processes of angular momentum loss in a stellar wind, which has
traditionally been modelled on the basis of an aligned dipole field
where most of the open (wind-bearing) field lines emerge from near the
stellar pole.  It is also relevant to studies of disk-magnetosphere
interaction and channeled accretion flows in young stars.

The purpose of this paper is to determine if there are any
observational tests that could distinguish between these three
different models for the nature of the polar field.  We do this by
modelling the effect of these different models on the structure of the
coronal magnetic field and the nature of its X-ray emission.

%----------------------------------------------------------------------
\begin{figure*}
  \def\subfigtopskip{4pt}
  \def\subfigbottomskip{4pt}
  \def\subfigcapskip{2pt}
  \centering
  \begin{tabular}{cc}
    \subfigure[]{
    \psfig{figure=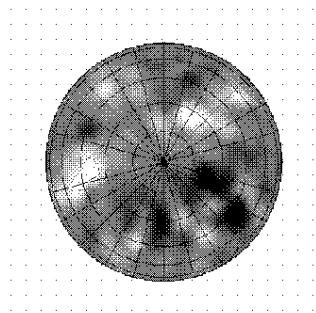,width=8cm}
    } &
    \subfigure[]{
    \psfig{figure=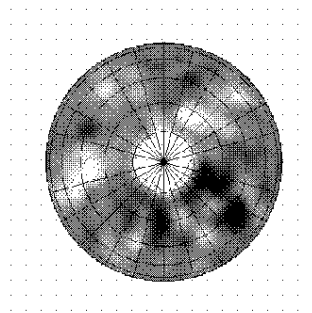,width=8cm}
    } \\
    \subfigure[]{
    \psfig{figure=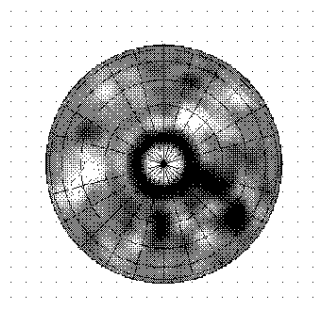,width=8cm}
    } &
    \subfigure[]{
    \psfig{figure=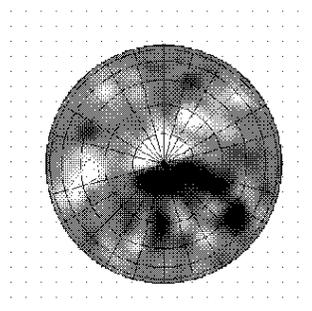,width=8cm}
    } \\

  \end{tabular}

 \caption{Surface maps of the magnetic field for AB Dor for Dec 1995 shown from
a viewpoint looking down on the rotation pole. Fig 1a shows the surface magnetic
field as it is observed with no field added to the polar region.  Fig 1b has a
large unipolar spot, Fig 1c, concentric rings of opposite polarity and fig 1d,
the bipolar configuration. The strength of the field added to each model is 1KG.
}

  \label{polarview}

\end{figure*}
%----------------------------------------------------------------------
\section{AB Doradus}

We take as our example star the young rapid rotator AB Doradus which
has been studied in depth for the last decade or so.  Surface brightness maps
have been collected annually since 1988
\citep{1994MNRAS.269..814C,1995MNRAS.277.1145U,1995MNRAS.275..534C}.
In the last seven years Zeeman-Doppler maps of the surface magnetic
field have been secured on an annual basis.  Long term studies of
the star's X-ray variability have shown a small rotational modulation
(5-13 \%) and a large emission measure of $10^{52-53}$cm$^{-3}$
\citep{1993A&A...278..467V}.  This implies that AB Dor has either a very
extended or very dense corona.  AB Dor has an inclination of
60$^{\circ}$ and therefore much information about the lower hemisphere
is lost.  In the upper hemisphere however, the pole remains in view
constantly and X-ray emission from high latitudes would therefore
suffer little rotational modulation.  BeppoSax observed a flare
\citep{2000A&A...356..627M} that showed no rotational modulation over
a period greater than the rotation period of the star.  Modelling of
the flare decay showed that the flaring loops must be small and hence
must be situated at latitudes above 60$^\circ$ in order to remain in view.

Complementary to these observations are spectroscopic studies of the
star's X-ray emission from which the thermal structure, abundance
stratification, and densities of the corona can be investigated.  Recent studies
with XMM showed coronal densities to be extremely high ranging from
$3\times 10^{10}$cm$^{-3}$ \citep{2001A&A...365L.336G} to
$10^{12}$cm$^{-3}$.  This high density immediately suggests that the emitting
loops would be relatively small and compact.  If this is the case, to explain
the lack of rotational modulation, these loops would have to be located at
latitudes high enough to remain in view as the star rotates.  High latitude
magnetic loops are consistent with Doppler images of AB Dor where we can see
dark polar spots along with spots at lower latitudes.  Zeeman-Doppler images
show magnetic flux to be present all over the surface of AB Dor apart from at
the poles where the surface brightness is so low that the Zeeman signal is
suppressed \citep{1997MNRAS.291....1D}.

Any model applied for the field at the pole will have to explain the
lack of rotational modulation in the X-ray emission as well as being
consistent with the observations of large prominences that form
preferentially at or just beyond the Keplerian co-rotation radius,
which for AB Dor is at 2.7 stellar radii from the rotation axis
\citep{1989MNRAS.236...57C,1989MNRAS.238..657C}.  As many as six
prominences may be present at any one time in the observable
hemisphere of the star.  Since they co-rotate with the star, they must
be held in place by the coronal magnetic field.  This suggests that at
least some fraction of the coronal field is in the form of closed
loops even at these large distances.

The structure of the large scale coronal magnetic field of AB Dor was
investigated by \citet{2002MNRAS.333..339J}.  They took the surface
magnetic field determined from Zeeman-Doppler imaging and extrapolated
it into the corona, assuming it to be potential.  They found that the
closed field regions of the corona extended over the polar regions.
By filling the corona with isothermal plasma in hydrostatic
equilibrium, they determined the spatial distribution of the coronal X-ray
emission \citep{2002MNRAS.336.1364J}.  Much of this emission came from
high-latitude regions where it was never rotationally self-eclipsed.
This naturally gave a low rotational modulation in X-rays, similar to
the observed value.  The derived magnitude of the emission measure and
the emission-measure weighted mean density were also consistent with
observations. Jardine {\em et al}also explored the effect on the X-ray emission
of adding in a global dipole and determined that because this would cause
the polar regions to be open (and hence dark in X-rays) this could not
be reconciled with the BeppoSAX flare observations that suggest the
presence of closed loops at high latitudes
\citep{2000A&A...356..627M}.

In this paper, we consider the impact of some alternative models that
may explain the dark polar caps.  Rather than a global dipole which
extends over the whole surface, we consider a single unipolar spot at
the pole which covers a limited surface area.  We also consider two
possibilities for mixed-polarity regions at the pole: either a single
bipole or concentric rings of alternating polarity.  These configurations
were added to the poles of the 1995-1996 mixed map of AB Dor.  In each
case the global field was extrapolated and the positions of stable points
which are possible prominence formation sites and the X-ray emission
were determined.  The effects on these properties were examined for
different strengths of polar field ranging from 500G to 2000G. Below this upper
limit, polar fields would escape detection. Although the polar areas on AB Dor
are dark and the Zeeman signal is thus suppressed,
\citet{1997MNRAS.291....1D}have shown that if the polar region
was to have a field in excess of a few thousand Gauss there would be some signal
of the polar field detected by Zeeman-Doppler imaging.

%----------------------------------------------------------------------
\begin{figure*}

  \begin{tabular}{cc}
  \subfigure[]{
   \psfig{figure=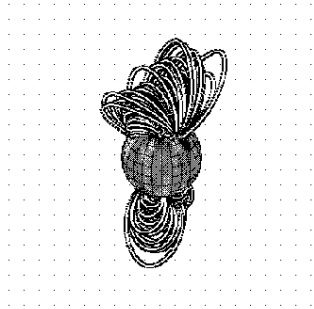,width=8cm}
   } &
   \subfigure[]{
   \psfig{figure=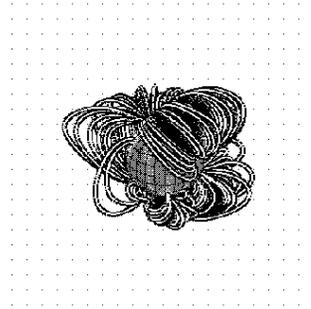,width=8cm}
   } \\
   \subfigure[]{
   \psfig{figure=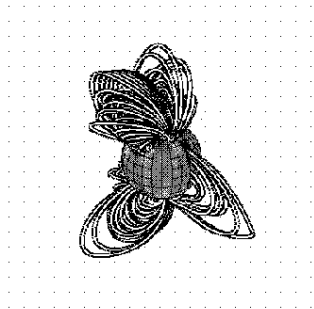,width=8cm}
   } &
   \subfigure[]{
   \psfig{figure=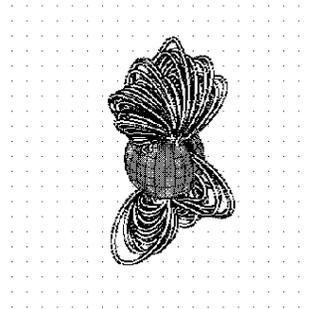,width=8cm}
   } \\

  \end{tabular}

 \caption{Closed field emerging only from the polar regions where we have added
flux. Strong similarities can be seen between the original image, concentric
rings and bipole models. In these cases the field lines from the polar region
connect primarily to the surface at high latitudes. Only in the case of a
unipolar spot (b) do the field lines connect to the rest of the stellar
surface.}

  \label{mixed0_{closed}}

\end{figure*}
%----------------------------------------------------------------------

%----------------------------------------------------------------------
\begin{figure*}

  \begin{tabular}{cc}
   \subfigure[]{
   \psfig{figure=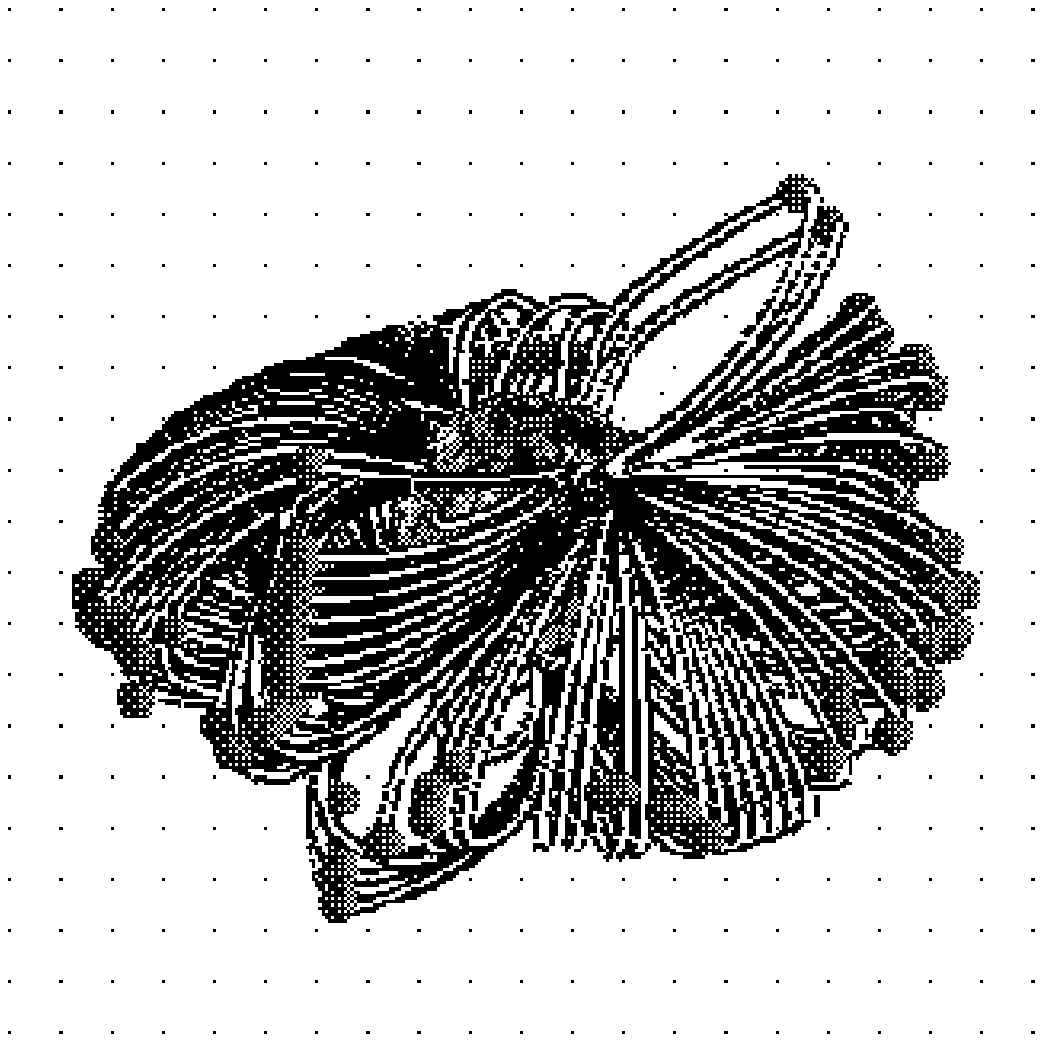,width=8cm}
   } &
   \subfigure[]{
   \psfig{figure=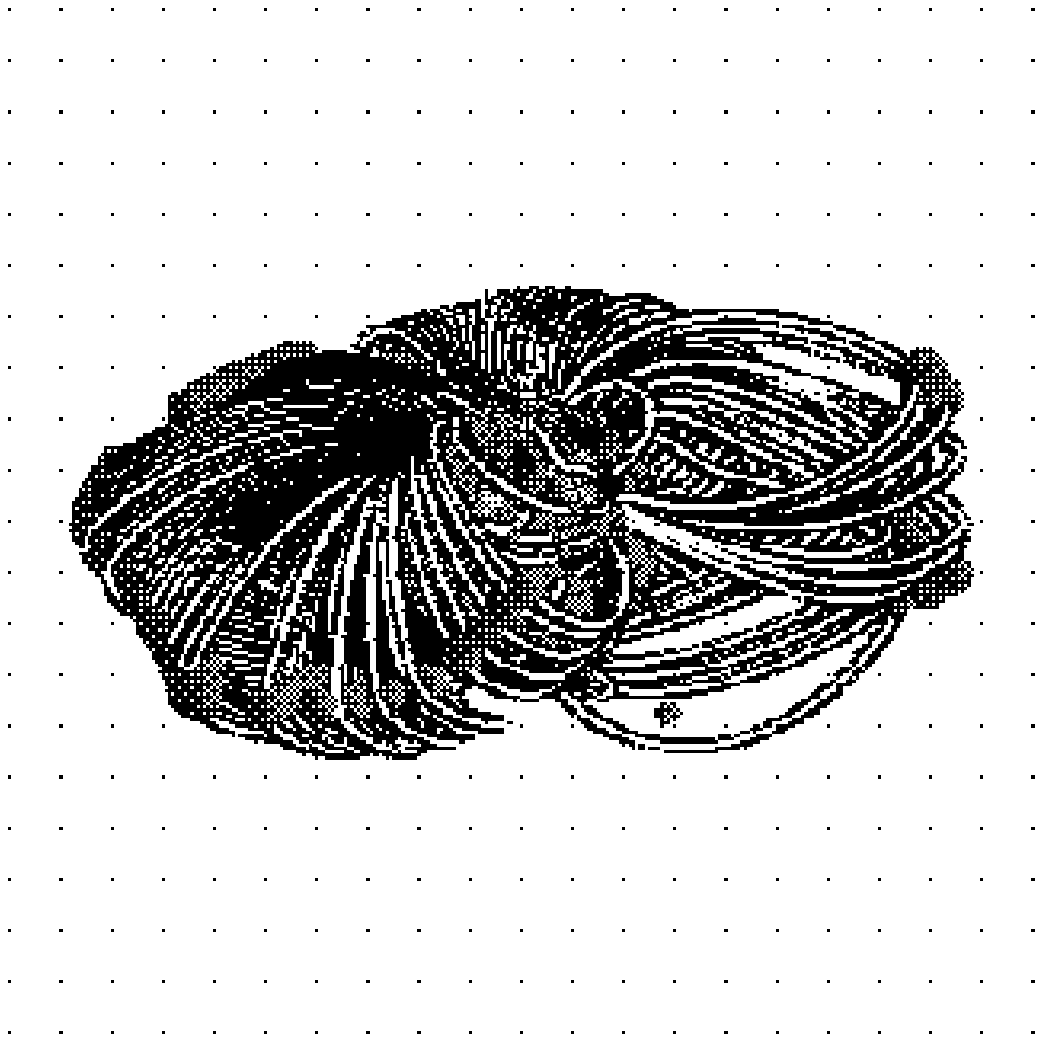,width=8cm}
   } \\
   \subfigure[]{
   \psfig{figure=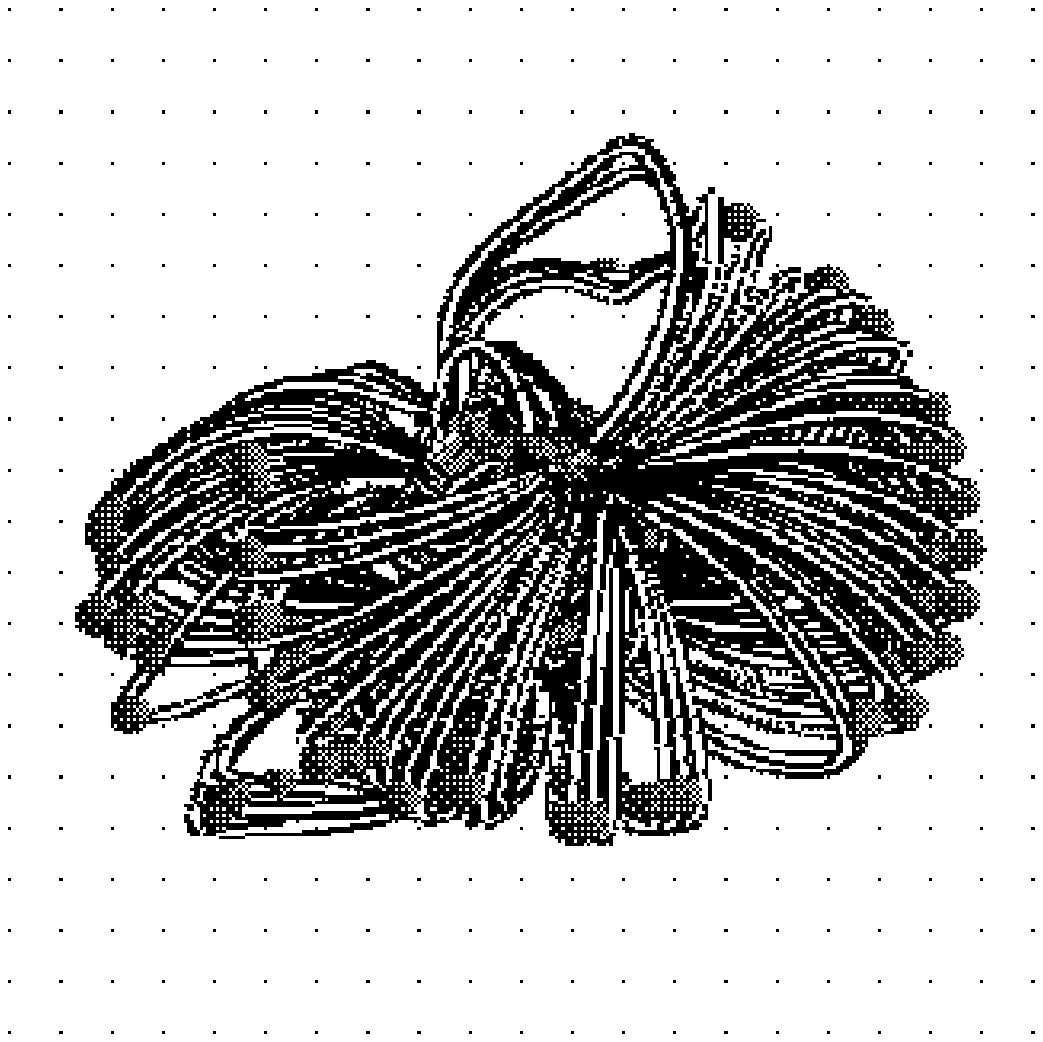,width=8cm}
   } &
   \subfigure[]{
   \psfig{figure=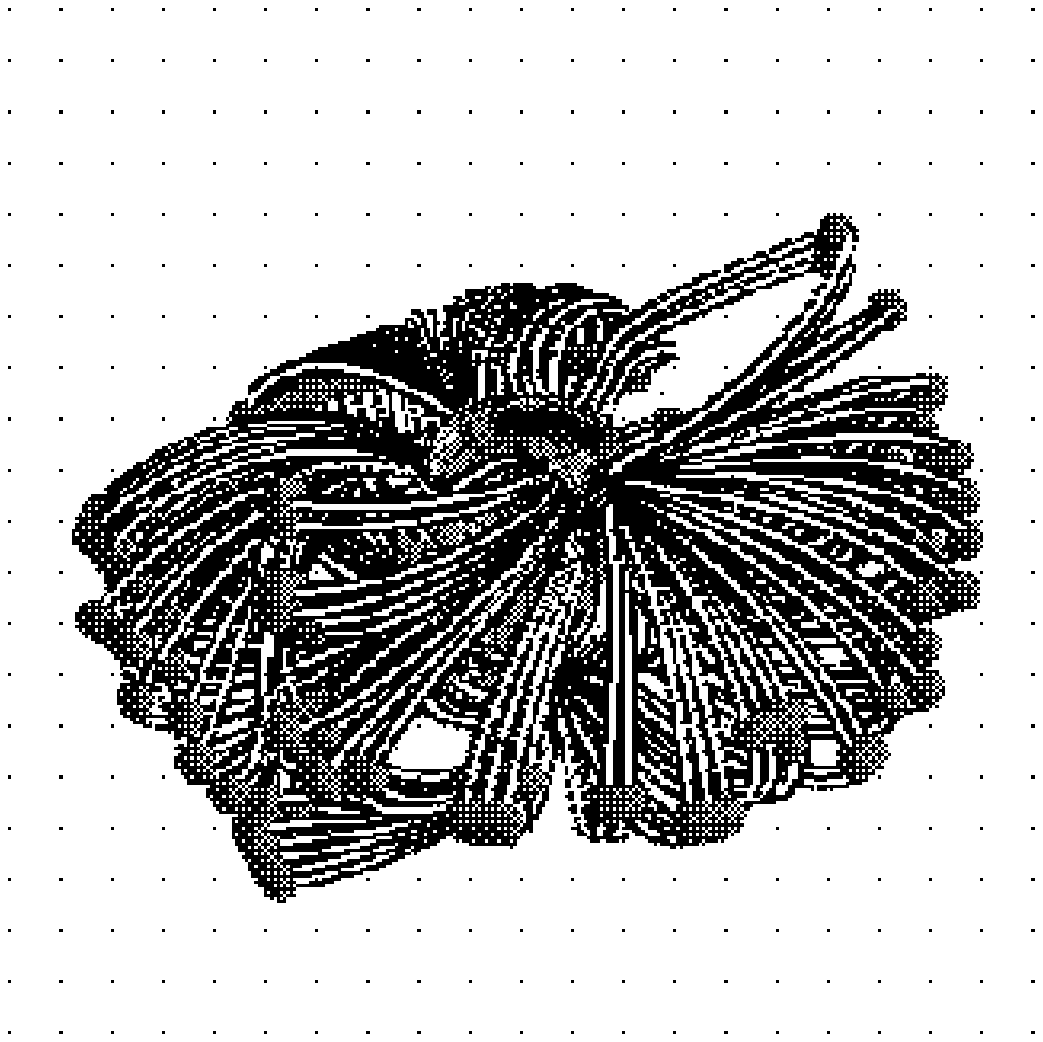,width=8cm}
   } \\

  \end{tabular}

 \caption{Surfaces with closed field lines show possible locations of
prominences.At each of these points the effective gravity along the field line
is zero and the point lies at a potential minima. moving from top left to bottom
right, polar conconfigurations considered are: nothing at the pole(a), global
dipole(b),
concentric rings(c), bipole at the pole(d). Each polar cap added has field
strengths of
1kG used.}

  \label{proms}

\end{figure*}
%----------------------------------------------------------------------

%----------------------------------------------------------------------
\begin{figure*}

  \begin{tabular}{cc}
   \subfigure[]{
   \psfig{figure=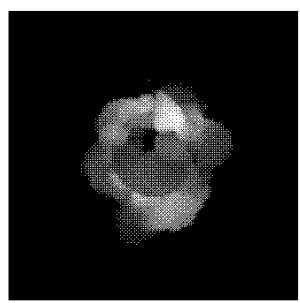,width=7.5cm}
   } &
   \subfigure[]{
   \psfig{figure=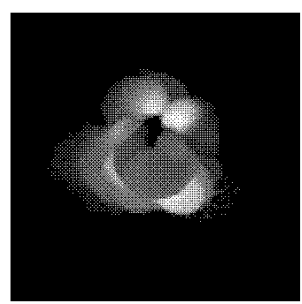,width=7.5cm}
   } \\
   \subfigure[]{
   \psfig{figure=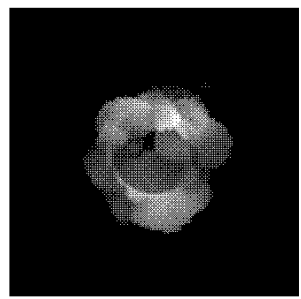,width=7.5cm}
   } &
   \subfigure[]{
   \psfig{figure=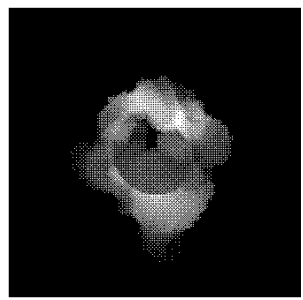,width=7.5cm}
   } \\

  \end{tabular}

 \caption{X-ray images of the stellar corona for each model. Same layout as
previous
figures. The X-ray emissions are are all calculated for a coronal
temperature T=$10^7$K. }

  \label{xray}

\end{figure*}
%----------------------------------------------------------------------

%----------------------------------------------------------------------
\begin{figure*}

   \begin{tabular}{cc}
   \subfigure[]{
   \psfig{figure=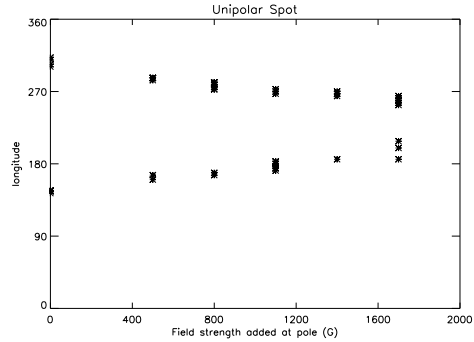,width=6.5cm}
   } \\
   \subfigure[]{
   \psfig{figure=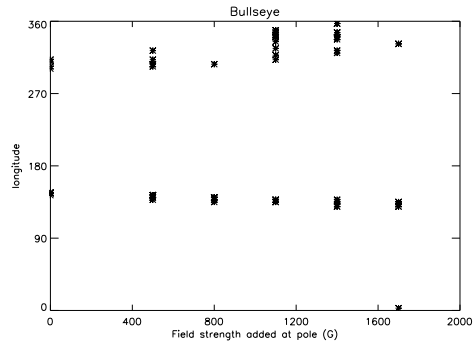,width=6.5cm}
   } \\
   \subfigure[]{
   \psfig{figure=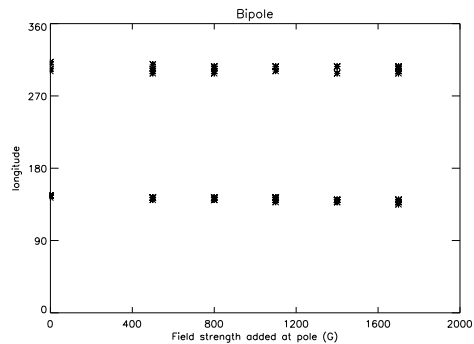,width=6.5cm}
   } \\
   \end{tabular}

 \caption{The graphs above shows the longitude for stable points where possible
prominces
may form and would pass in front of the star so they may be detected. The
results here are
fairly conclusive showing that all three models have roughly two regions of
longitude
where prominences could be detected.  }

  \label{longs}

\end{figure*}
%----------------------------------------------------------------------
%----------------------------------------------------------------------
\begin{figure*}

   \psfig{figure=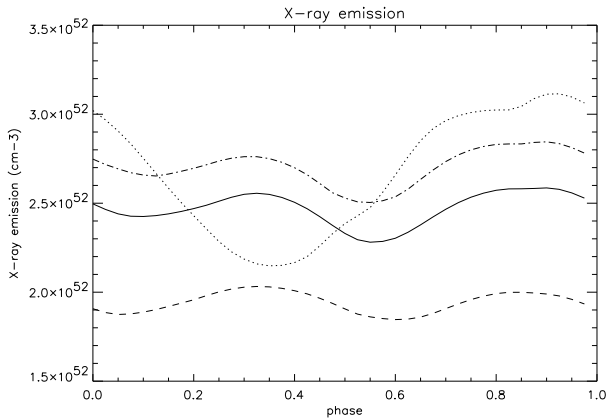,width=8cm}

 \caption{Graph showing the X-ray emission measure as it varies with phase for
each
model. All models here do not differ too much in magnitude and certainly the
original
data(solid line), concentric rings(dashed line) and Bipole(dash-dots) show a
remarkably
similar curve in emission over a  whole cycle. This is also reflected in their
rotational modulation which shows little variation between these three models
and fits
into the range of values predicted by observations of 5-13\%
\citep{1997A&A...320..831K}.
The Unipolar spot model (dotted line) shows a much higher rotational modulation
of
31.52\%.}

  \label{xraymod}

\end{figure*}
%----------------------------------------------------------------------
%----------------------------------------------------------------------
\begin{figure*}

   \psfig{figure=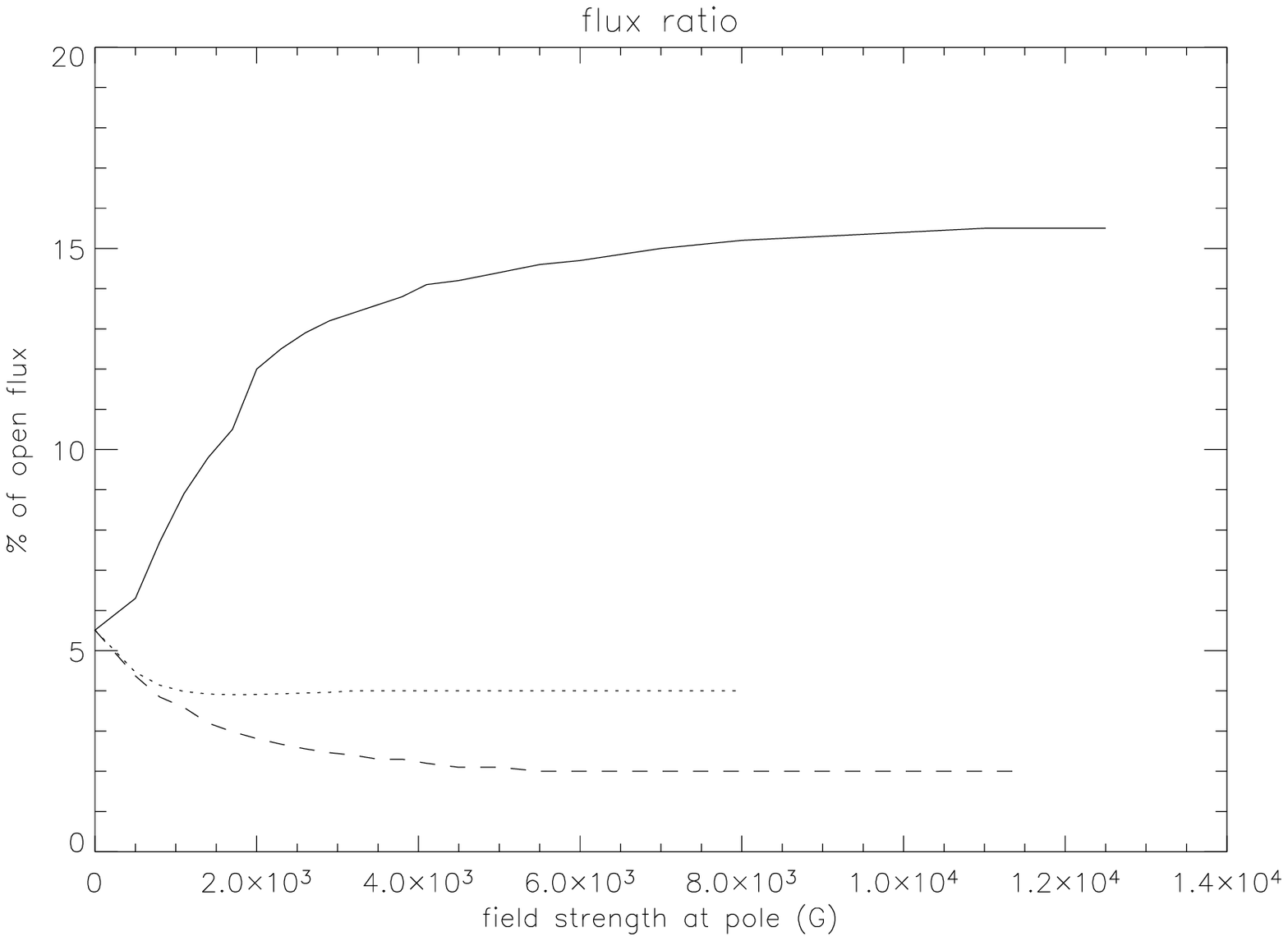,width=8cm}

 \caption{The graph shows the change in the amount of open magnetic flux as we
increase
the field strength at the poles of the various models. Any flux that reaches the
source
surface is considered open hence we calculate the amount of open flux as a ratio
of total
flux emerging at the stellar surface:
%\begin{equation}
$\sum_{A}|B_{r}(R_{s},\theta,\phi)| / \sum_{A}|B_{r}(R_{*}, \theta,\phi)|$
%\end{equation}
The solid line shows the change for the unipolar spot model, the dotted line
represents
the concentric rings model and the dashed line the bipolar model. Again the
similarities
between the concentric rings and bipole models is easily seen, both slowly
decreasing in
open flux as we increase the strength of the polar field. The unipolar spot
model on the
other hand shows a reasonable growth in open flux with increasing field strength
at the poles. It should be noted that although we have shown the flux ratios for magnetic 
fields up to 12kG these fields are much too strong to be considered non-dectable and are 
only shown to display the saturation each of the models has on the open flux.}

  \label{mixed0_open}

\end{figure*}
%----------------------------------------------------------------------

\section{Extrapolating the coronal field}

The extrapolation of the field is the same as used in Jardine, Collier
Cameron \& Donati (2002). Since the method is the same we refer to the
previous mentioned paper for a detailed description of the calculations
involved. We write the magnetic field $\bvec{B}$ in term of a flux
function $\Psi$ such that $\bvec{B} = -\bvec{\nabla} \Psi$ and the
condition that the field is potential ($\bvec{\nabla}\times\bvec{B} =0$) is
satisfied automatically. The condition that the field is divergence-free
then reduces to Laplace's equation $\bvec{\nabla}^2 \Psi=0$. A solution in
terms of spherical harmonics can then be found:
\begin{equation}
\Psi=\sum_{l=1}^{N}\sum_{m=-l}^{l} [a_{lm}r^l + b_{lm}r^{-(l+1)}]
P_{lm}(\theta) e^{i m \phi}, \label{psifull}
\end{equation}
where the associated Legendre functions are denoted by $P_{lm}$.
The coefficients $a_{lm}$ and $b_{lm}$ are determined by imposing the
radial field at the surface from the Zeeman-Doppler maps and assuming that
at some height $R_s$ above the surface the field becomes radial. Due to the
presence of large slingshot prominences observed around the co-rotation
radius which lies at $2.7R_\star$ from the rotation axis, we believe that much
of the corona is closed out to those heights and so we set the value of
$R_s$ to $3.4R_\star$.  In order to calculate the field we used a code
originally developed by \citet{1998ApJ...501..866V}.

\section{calculating the position of prominences}
The numerical method for determining the locations of stable potential minima
along field lines is discussed in detail by \citet{2002MNRAS.330..160P} and will
only be given a brief review here. For a possible prominence formation site to
exist the component of effective gravity along the magnetic field must be zero,
i.e. $\bvec{B}.\bvec{g_{eff}}=0$. For this position of equilibrium to be stable
we also require it to be a potential minimum. Thus the equilibrium must satisfy
the condition
 \begin{equation}
  (\bvec{B}.{\nabla})(\bvec{B}.\bvec{g_{eff}})<0.
 \end{equation}
Such stable points may be favoured positions for prominence formation as any gas
condensing at this point would neither fall towards the surface nor be
centrifugally expelled.  The fact that prominences on AB Dor are observed around the
co-rotation radius suggests that such minima in the gravitational-centrifugal potential do
have a role to play in determining where prominences form. The existence of such stable
points does not however guarantee prominence formation. In fact, at distances as far out
as the co-rotation radius, the curvature of the field lines would need to be much smaller
in order to overcome the centrifugal acceleration acting on a prominence. Prominences are
observed as transient $H_{\alpha}$ absorption features and so can only be seen when they
occult the stellar disk. We therefore select out of all possible stable points those that
transit in front of the disk. Hence due to the stellar inclination any points
that will transit the disk must satisfy the condition
 \begin{equation}
    R\cos(\theta + 60)<1
   \end{equation}
where $R$ is the radial height of the stable point and $\theta$ is the latitude.
If a star has a purely dipolar field, all such stable points would lie in the
equatorial plane and so would be unobservable \citep{2001MNRAS.324..201J}. We
anticipate with our unipolar spot model that as the field strength in the polar
cap is increased, the polar field will eventually dominate over the low-latitude
field and the global topology will resemble that of a dipole. In this case no
stable points in the corona would be observable. For each of our polar field
models, we have therefore investigated the number of stable points that
could be observed. Fig.5 shows the longitudes of these stable points.

\section{X-ray emission} The X-ray emission was determined from the closed
field regions. First of all the pressure structure of the corona was calculated
assuming
it to be isothermal and in hydrostatic equilibrium. Thus the pressure at any
point is
\begin{equation}
p = p_{0}e^{\int g_{s} ds}
\end{equation}
where $g_{s} =( {\bf g.B})/|{\bf B}|$ is the component of gravity along the
field and
\begin{equation}
{\bf g}(r,\theta) = \left( -GM_{\star}/r^{2} +
                     \omega^{2}r\sin^{2}\theta,
		     \omega^{2}r\sin\theta\cos\theta
             \right),
\end{equation}
with $\omega$ the stellar rotation rate.  The plasma pressure is scaled to the
magnetic pressure at the loop footpoints i.e. $p\propto {B_{0}}^2$ where the constant of
proportionality is chosen to match the observed emisssion measure.  As we move out to
larger heights from the star the gas pressure is greater than the magnetic pressure and
thus forces field lines to open up. We have included this effect in our model by setting
the plasma pressure to zero wherever it is greater than the magnetic pressure i.e. where
$\beta>1$.  From the pressure, the corona's density could be calculated assuming it to be
an ideal gas.  Using a Monte Carlo radiative transfer code we can then determine the X-ray
emission.

\section{Results}
The magnetic fields added to the star are depicted in Fig.1 as is the
original surface field map.  All three additional fields have a
strength of 1kG. This is a reasonable estimate
for the field at the pole as it is strong enough to suppress
convection yet not so strong as to be detectable in the star's
Zeeman-Doppler images.

As can be seen in Fig.2, where we have shown only those field lines
that are anchored at points within the polar region in question, the
unipolar spot model has the greatest effect on the global field
topology.  Closed loops from the polar
region are forced down to lower latitudes due to the presence of
strong open field at the pole.  This is unlike the other two models with
the concentric rings and bipole, which show a strikingly similar
appearance to the original data set of having large closed loops
crossing directly over the pole.  In these models there is
almost no open field emerging from the polar region.  The concentric
rings and bipole models show greater consistancy in the lack of rotational
modulation of the flaring X-ray emission seen in BeppoSAX observation.

The difference in the topology of the field that is added at the pole
can be seen in Fig (\ref{mixed0_open}) which shows the fraction of the
surface flux that is open.  In the case of a single unipolar spot
added to the pole, this fraction increases as the strength of the
polar field is increased, reaching a maximum value of $15\%$.
This is less than the corresponding value of $44\%$ for a dipolar
field with the same source surface imposed \citep{2002MNRAS.333..339J} but
still represents a significant factor.  The open field regions extend down to a
latitude of 75$^\circ$ with the result that much of the polar regions would be
dark in X-rays.  For the other two types of polar field, however, most of the
polar field lines are closed (see Fig (\ref{mixed0_{closed}})) and so
very few of them contribute to the flux of open field.  In this case,
the fraction of open field decreases as the polar field strength increases.
This is due to the field at the pole being able to connect locally.

Positions of possible prominences are less affected by the
addition of a polar field.  Again the unipolar spot has the greatest
effect forcing the stable points toward the equatorial plane
of the star (Fig.3).  Many of these points which lie in or near the equatorial
plane cannot be observed crossing the star along our line of sight.
Observations of prominences are possible only if the prominences
transit our line of sight to the star causing absorption dips in
the H$\alpha$ profile.  Strangely however the number of stable points
for the dipole model is not that much less than any of the other
models, although there does seem to be some convergence of longitudes
as we increase the field strength at the pole for this model.  The
concentric rings and bipole models reproduce the
same pattern of sites as the original image.  In all three models we find that
the stable points are grouped into two regions where arcades of predominantly
east-west field span a range of longitudes.The imposition of the polar field
disturbs the summits of these structures, pushing them toward longitude
220$^\circ$ in the unipolar case, and away from longitude 220$^\circ$
in the concentric case.

Clearly, none of these models can fully explain the number of
prominences that are observed (up to six in the observable hemisphere
at any one time).  One possibility is that these stable points are not
reliable indicators of prominence formation sites.  Alternatively, as
suggested by \citet{2001ApJ...551.1099S} it could be that the prominences are
formed in the sheared field produced when differential rotation
stretches out the unipolar field of the dark polar cap to produce an
azimuthal field.  Such fields have been observed on AB Dor for some
time \citep{1997MNRAS.291....1D}.  Recently, \citet{2002ApJ...575.1078H} have
fitted non-potential magnetic fields to the observed Stokes profiles.  They have
shown that the currents (which mark the regions where the field departs from its
lowest-energy state) are confined to the high latitude regions at the
edge of the dark polar cap.  The positions of the stable points in
these field extrapolations, however, were not significantly different
from those determined for a potential field.

\citet{2002MNRAS.330..160P} performed a potential field extrapolation from a
Zeeman-Doppler image of AB Dor with a dipolar field added into the
map.  They showed that differential rotation acting on such a field
would produce a high latitude ring of azimuthal field similar to that
observed \citep{1997MNRAS.291....1D}.  It is possible that the field produced by the flux
emergence model of \citet{2001ApJ...551.1099S}would have a similar effect. Here the
differential rotation between the concentric rings of opposite polarity would shear the
field crossing this boundary producing an azimuthal field.
A third possibility is that, as suggested by \citet{2002ApJ...575.1078H}, the closed
corona is in fact confined to within $1.6 R_{\star}$ and the prominences are confined in
the very cusps of helmet streamers, much further out than the rest of the corona.

The other observable quantity is the X-ray emission.  This was
calculated for each model with polar fields of 1kG. Looking at the
X-ray images of the stellar corona (Fig.4) for each model it is clear
that the original image, the concentric rings and bipole models all show a
strong similarity with a predominant X-ray emission from the polar
region.  The unipolar spot does show some emission from the polar
region although not nearly as much structure as the other models.  The
average emission measure did not vary significantly between the
models, ranging from $1.9\times10^{52}$cm$^{-3}$ for the concentric
rings to $2.73\times10^{52}$cm$^{-3}$ for the bipole.  The unipolar
spot model has the highest emission measure with a value of
$3\times10^{52}$cm$^{-3}$ but it has such a large rotational
modulation (31.5\%) that its average emission lies at
$2.61\times10^{52}$cm$^{-3}$.  The other models all show a level of
rotational modulation (9 to 11\%) that fits the values predicted by
observations of 5 to 13\% \citep{1997A&A...320..831K}.  Something else
worth noting here is that the models, excluding the unipolar spot, all
showed very similar X-ray light curves.  This again strongly suggests that
the presence of a mixed-polarity field at the pole has little effect on
the overall structure of the star.

\section{Conclusions}
Using Zeeman-Doppler images of the surface magnetic field of the
young, rapidly rotating dwarf AB Dor, we have investigated the effect
on the coronal field structure of the three most popular models for
the polar fields of young stars.  We find that the addition of a
mixed-polarity region at the pole results in more small-scale local
flux tubes at the pole, but has little effect on the large-scale field
structure.  Consequently, observations of the magnitude or rotational
modulation of the X-ray emission are not sufficient to discriminate
between these types of models.  Observations of the ``slingshot
prominences'' trapped in the coronae of these stars do not provide an
observational discriminant either. We have determined possible
prominence-bearing  structures by calculating the sites of stable mechanical
equilibrium. All three models showed no more than 2 structures
that can transit the stellar disc, but the observations show condensations at
all longitudes. Due to the nature of alternating magnetic polarity
at the surface of the star it is thought that a helmet streamer model could
provide us with more regions of longitude for possible prominence formation
sites where each prominence site would lie above and between these regions of
opposite polarity \citep{2001JGR...10625165L}.

The large-scale field structure is affected rather more by the addition
of a unipolar region at the pole. This forces more of the
polar field lines to be open (and hence dark in X-rays). This gives a
larger rotational modulation as the closed-field regions that are
bright in X-rays are now at lower latitudes and so are subject to
rotational self-eclipse.

A polar spot extending down to latitude 75$^\circ$ gives a rotational modulation 
of 31.5\%, greater than the observed value of 5-13\%
\citep{1997A&A...320..831K}. It also forces up to 15\% of the flux that emerges
through the surface to be open, compared to only 5\% when no field is added at 
the pole. This has potential implications for the angular momentum lost in the 
stellar wind, not only because it changes the amount of open flux, but also 
because it affects the distribution in latitudes. The addition of a polar spot 
introduces many more high-latitude open field lines. The winds from high 
latitude regions remove significantly less angular momentum from the star 
because of the reduced lever arm of the fieldlines. \citet{1992A&A...264L..13S} 
showed that this may explain the apparent slow-down of the angular momentum loss 
in young rapid rotators.

The change in the global field structure brought about by adding a polar spot 
may also be relevant to studies of disk-magnetosphere coupling in young stars. 
The addition of a polar spot means that the largest scale field lines ( the ones 
that would intersect a disk ) originate from the poles. In this case, any 
material accreting from a disk would reach the stellar surface at the poles. In 
the absence of a polar spot ( or in the case where the polar field was of mixed 
polarity) the accretion process would be more likely to be in the form of 
discrete accretion funnels, which would intersect the stellar surface in 
low-latitude 'hot spots'.

While the nature of the polar field - whether single or mixed polarity - clearly 
makes a difference to the global field structure, the available observations of 
the X-ray emission measure and the prominence distribution are not sufficient to
discriminate between types of polar field. We can however rule out the posibility of the 
unipolar spot model because the results show the modulation amplitude in the X-ray 
emission to be much too large. Only the degree of modulation of the X-ray emission can be
used, but this requires long-term monitoring to eliminate the effects of short-term
variablity in flares. Phase-resolved observations of line shifts that indicate the
presence of a stellar wind would perhaps be a good proof of the field structure, but it
may be that it will only be with the advent of Zeeman-Doppler imaging in molecular lines
that we will be able to determine the nature of the polar fields.

\section{References}
\bibliographystyle{test}
\bibliography{literature,journals}

\end{document}